\documentclass[11pt]{article}
\usepackage[bf,hang]{subfigure}
\usepackage{graphicx}
\usepackage[bf,hang]{caption}

\title{\Large Avoiding Braess' Paradox through Collective Intelligence}

\author{Kagan Tumer\\
NASA Ames Research Center\\
Moffett Field, CA 94035\\
kagan@ptolemy.arc.nasa.gov\\
\and
David H. Wolpert\\
NASA Ames Research Center\\
Moffett Field, CA 94035\\
dhw@ptolemy.arc.nasa.gov \\ 
\and
Tech Rep No. NASA-ARC-IC-1999-124}

\begin{document}

\maketitle

\begin{abstract} 
In an Ideal Shortest Path Algorithm (ISPA), at each moment each
router in a network sends all of its traffic down the path that
will incur the lowest cost to that traffic. In the limit of an
infinitesimally small
amount of traffic for a particular router, its routing that traffic
via an ISPA is optimal, as far as cost incurred by that traffic
is concerned. We demonstrate though that in many cases, due to
the side-effects of one router's actions on another routers
performance, having routers use ISPA's is suboptimal as far as
global aggregate cost is concerned, even when only used to
route infinitesimally small amounts of traffic.  As a particular
example of this we present an instance of Braess' paradox for ISPA's,
in which adding new links to a network decreases overall
throughput. We also  demonstrate that load-balancing, in which the
routing decisions are made to optimize the global cost incurred by all
traffic currently being routed, is suboptimal as far as
global cost averaged across time is concerned. This is also due
to ``side-effects", in this case of current routing decision on future
traffic.
The theory of COllective INtelligence (COIN) is concerned precisely
with the issue of avoiding such deleterious side-effects. We present
key concepts from that theory and use them to derive an idealized
algorithm whose performance is better than that of the
ISPA, even in the infinitesimal limit. We present
experiments verifying this, and also showing that a  machine-learning-based
version of this COIN algorithm in which costs are only imprecisely
estimated (a version potentially applicable in the real
world) also outperforms the ISPA, despite having access to less
information than does the ISPA. In particular, this COIN algorithm
avoids Braess' paradox.

\end{abstract}

\nocite{hesn98}

\nocite{boli94}

\nocite{sudr97}

\nocite{chye96}

\nocite{mami98}

\section{INTRODUCTION} 

The problem of how to control routing across a network underlies a
vast array of real-world problems including internet routing,
voice/video communication, traffic flows, etc.  In its general form,
the problem is how to optimize the flow of certain entities (e.g.,
information packets, cars) from sources to destinations across a
network of routing nodes.  Here we are concerned with the version of
the problem in which ``optimization'' consists of minimizing aggregate
cost incurred by the entities flowing to their destinations. To ground
the discussion, we will consider the case where the entities being
routed are packets.

Currently, many real-world network routing solutions to this
particular problem are based on the Shortest Path Algorithm (SPA), in
which each routing node in the network maintains estimates of the
``shortest paths'' (i.e., minimal total incurred costs) from it to
each of its destinations and at each moment satisfies any routing
requests by sending all its packets down that shortest path.  Many
algorithms exist for efficiently computing the shortest path in the
case where the costs for traversing each component of every path at
any given time are known. In particular, there exist many such
algorithms that can be applied when node-to-node path-cost
communication is available and the costs for traversing each component
are unvarying in time (e.g., Dijkstra's Algorithm
\cite{ahma93,bega92,depa84,dijk59}. Real-world SPA's apply such
algorithms to estimated costs for traversing each component of every
path to generate their estimated shortest paths.

Consider the case where for all paths from a particular node to a
particular destination, the costs that would be incurred by that
node's routing all its current traffic along that path is known
exactly to that node (the information being stored in that router's
``routing table''). Clearly if a non-infinitesimal amount of traffic
is being routed by our node, then in general its sending all that
traffic down a single path will not result in minimal cost incurred by
that traffic, no matter how that single path is chosen. However if it
must choose a single path for all its traffic, then tautologically
the SPA chooses the best such path. Accordingly, in the limit of
routing an infinitesimally small amount of traffic, with all other
nodes' strategies being a ``background'', such a router's running SPA
is the optimal (least aggregate incurred cost) routing strategy {\it
for that particular routing node considered individually}.

One might hope that more generally, if the node must allot all of its
traffic to a single path, then its choosing that path via the SPA
would be the $globally$ optimal choice of a single path, at least in the
limit of infinitesimally little traffic.  This is not the case though,
because in using the SPA the node is not concerned with the
deleterious side-effects of its actions on the costs to other
nodes~\cite{kola97,wotu99a}. In the extreme case, as elaborated below,
if all nodes were to try to
minimize their personal costs via SPA's, then the nodes would actually {\it
all} receive higher cost than would be the case under an alternative
set of strategies. This is an instance of the famous Tragedy Of the
Commons (TOC)~\cite{hard68}.

Deleterious side-effects need not be restricted to extend over space;
they can also extend over time. Indeed, consider the algorithm of
having all routers at a given moment make routing decisions that
optimize global cost incurred by the traffic {\it currently being
routed}, an algorithm often called ``load-balancing'' (LB). By
definition, LB avoids the deleterious side-effects over space that can
result in the TOC for the costs incurred by the traffic currently
being routed.  However, due to side-effects over time, even
conventional LB is often suboptimal as far as global cost averaged
across time is concerned. Intuitively, one would have to use
``load-balancing over time" to ensure truly optimal performance.

In this paper we are concerned with how to address these kinds of
deleterious side-effects, and thereby improve
performance. In particular, we are interested in ways of doing this
that result in better performance than that of the ubiquitous SPA.

Now use of the SPA obviously provides no guarantees, even for personal
cost of the router using it, if the path-estimates of the nodes are
incorrect. Such inaccuracy is the rule rather than the exception in
many practical applications. Typically those estimates will be in
error because node-to-node communication is not instantaneous, and
therefore routing tables may be based on out of date information.
More generally though, even if that communication were instantaneous,
the cost to traverse a component of the network may be different by
the time the packet arrives at that component.

In this paper we do not wish to investigate such topics, but rather to
highlight the issue of side-effects. Accordingly we ``rig the game''
in favor of the SPA by constructing our simulations so that the first
potential cause of routing table inaccuracy does not arise, and the
second is minimized. We do this in our experiments by using an {\em
Ideal} Shortest Path Algorithm (ISPA) which has direct access to the
shortest path at each moment. Note that this ISPA provides an
upper bound on the performance of any real-world SPA.

In general, even without side-effects, determining the optimal
solution to a flow problem (e.g., determining what the loads on each
link need to be to maximize throughput on a non-cooperative data
network) can be nontractable~\cite{ahma93,orro93a}.  Therefore, we
will concern ourselves with providing {\em good} solutions that avoid
the difficulties the ISPA has with side-effects. It is not our aim here
to present algorithms that find the best possible (``load-balanced
over time'') solution.

We will base our solutions on the concept of Collective Intelligence.
A ``COllective INtelligence'' (COIN) is any pair of a large,
distributed collection of interacting goal-driven computational
processes among which there is little to no centralized communication
or control, together with a `world utility' function that rates the
possible dynamic histories of the collection~\cite{wotu99a,wotu99b}.
In this paper we are particularly concerned with computational
processes that use machine learning techniques (e.g., reinforcement
learning~\cite{kali96,suba98,sutt88,wada92}) to try to achieve their goal, 
conventionally
represented as maximizing an associated utility function. We consider
the central COIN design problem: {\em how, without any detailed
modeling of the overall system, can one set utility functions for the
individual components in a COIN to have the overall dynamics reliably
and robustly achieve large values of the provided world utility?} In
other words, how can we leverage an assumption that our learners are
individually fairly good at what they do? In a
routing context, this question reduces to what goals one ought to
provide to each router so that each router's greedily pursuing those
goals will maximize throughput (``incentive engineering''). For
reasons given above, we know that the answer to this question is not
provided by SPA's goals --- some new set of goals is needed.

In Section~\ref{sec:back} we discuss the SPA's deficiencies and in
particular their manifestations in Braess' paradox.  We also
demonstrate the suboptimality of load-balancing in that section. We
then present Collective Intelligence in Section~\ref{sec:coin},
discuss the routing model we will use in our experiments, and show how
the theory of COINs can be applied to that model to provide an
alternative to shortest path algorithms.  In Section~\ref{sec:sim} we
present simulation results with that model that demonstrate that in
networks running ISPA, the per packet costs can be as much as 32 \%
higher than in networks running algorithms based on COIN theory. In
particular, even though it only has access to imprecise estimates of
costs (a handicap that does not hold for ISPA), the COIN-based algorithm 
almost always avoids Braess' paradox,
in stark contrast to the ISPA. In that the cost incurred
with ISPA's is presumably a lower bound on that of an SPA not privy to
instantaneous communication, the implication is that COINs can
outperform such real-world SPA's.\footnote{A brief synopsis of the
COIN algorithm discussed here was presented in a space-constrained
article ~\cite{wotu99a}; this paper presents full details and applies the
algorithm to Braess' paradox as an illustration of the suboptimality
of the SPA.}

\section{Suboptimality of Shortest Path and Load-Balancing}
\label{sec:back}
In this section we first demonstrate the suboptimality of an SPA
when we have multiple nodes making simultaneous routing decisions,
where neither node  knows ahead of time the other's choice, and therefore
does not know ahead of time exactly what the costs will be. We then
demonstrate that such suboptimality can hold even when only one node is making a
decision, and it knows what decisions the others have previously
made. Next we present Braess' paradox, a particularly pointed instance
of these effects. (See ~\cite{bass92,coke90,coje97,kola99} for other discussion of Braess'
paradox in SPA routing.) We end by demonstrating the suboptimality of
conventional load-balancing when cost over time is what's of interest.

\subsection{SPA when multiple routers are simultaneously
making decisions}
\label{sec:ispa}
Perhaps the simplest example of how individual greed on the part of all
nodes can lead to their collective detriment occurs when two nodes
determine that their shortest path is through a shared link with a
limited capacity, while both have a second option that is slightly
less preferable. In such a case, their using the common link degrades
the performance of {\em both} parties, since due to limited capacity
the performance of that link will quickly fall below that of their
second option.

More precisely, consider the case where, given a load $x$, the shared
link has a cost given by $x^3$, and where each router has a second
option where the cost is given by $2x$. Acting alone, with a single
packet to send, they would both send that packet through the shared
link (cost of 1). However by both doing so, they incur a larger cost
(cost of 8) than if they had both used their second choice (cost of
4). Without knowing what each other will do ahead of time (information
not conventionally contained in routing tables), the nodes will
necessarily have mistaken cost estimates and therefore make incorrect
routing decisions. (Indeed, to have $all$ nodes know what each other
are doing ahead of time requires the use of game theory.) In this,
even in the limit of differentially small packets, use of SPA will
lead to a wrong routing decision.

\subsection{SPA when only one router is making a decision}
\label{sec:subopt}

\begin{figure} [htb]
\begin{picture}(100,140)(-100,-30)
\put(50,0){\circle*{10}}
  \put (50,-15) {\makebox(-1,1)[b]{$X$}}
     \put(50,0) {\line (-2,3){50}}
     \put(50,0) {\line (2,3){50}}
\put(150,0){\circle*{10}}
  \put (150,-15) {\makebox(-1,1)[b]{$Y$}}
     \put(150,0) {\line (-2,3){50}}
     \put(150,0) {\line (2,3){50}}
\put(0,75){\circle*{10}}
  \put (0,90) {\makebox(-1,1)[t]{$A$}}
\put(100,75){\circle*{10}}
  \put (100,90) {\makebox(-1,1)[t]{$B$}}
\put(200,75){\circle*{10}}
  \put (200,90) {\makebox(-1,1)[t]{$C$}}
\end{picture}
\caption{Independent decisions at the source}
\label{fig:simple}
\end{figure}
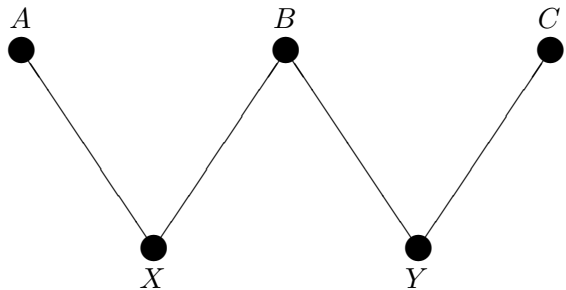

Consider the network shown in Figure~\ref{fig:simple}.  Two source
routers $X$ and $Y$ each send one packet at a time, with $X$ sending
to either intermediate router $A$ or $B$, and $Y$ sending to either
$B$ or $C$. This type of network may arise in many different
topologies as a subnetwork. Accordingly, difficulties associated with
this network can also apply to many more complex topologies.

Let $x_A$, $x_B$, $y_B$, and $y_C$, be the packet quantities at a
particular fixed time $t$, at $A$, $B$, or $C$, and originating from
$X$ or $Y$, as indicated.  At $t$, each source has one packet to
send. So each of our variables is binary, with $x_A + x_B = y_B + y_C =
1$. Have $V_i(z_i)$ be the cost, per packet, at the single instant
$t$, at router $i$, when the total number of packets at that instant
on that router is $z_i$. So the total cost incurred by all packets at
the time $t$, $G(\vec{x}, \vec{y})$, equals $x_A V_A(x_A) + (x_B +
y_B) V_B(x_B + y_B) + (y_C) V_C(y_C)$.

In an ISPA, $X$ chooses which of $x_A$ or $x_B$ = 1 so as to minimize
the cost {\it incurred by X's packet alone}, $g_X(\vec{x}) \equiv x_A
V_A(x_A) + x_B V_B(x_B + y_B)$. (Real-world SPA's typically try to
approximate this by having $X$ choose either $A$ or $B$ according to
whether $V_A(0)$ or $V_B(y_B)$ is smaller, where those two values can
be estimated via pings, for example.) In doing this the ISPA ignores the
$y_B V_B(x_B + y_B)$ term, i.e., it ignores the ``side effects'' of
$X$'s decision.

The right thing to do of course is instead have $X$ minimize
$G(\vec{x}, \vec{y})$, or more precisely, the components of
$G(\vec{x}, \vec{y})$ that depend on $X$.
Writing it out for this case, $X$ ought to act to minimize $x_A V_A(x_A)
+ (x_B + y_B) V_B(x_B + y_B)$. Due to the constraint that $x_A + x_B =
1$, this means sending down $A$ iff $V_A(1) < (y_B + 1) V_B(y_B + 1) -
y_B V_B(y_B)$, which differs from the ISPA result in that $X$ is
concerned with the full cost of going through router $B$, not just the
portion of that cost that its packet receives.

In the context of this example, this $G$-minimizing algorithm
constitutes ``load-balancing'' (LB). 
Note that so long as sgn$[V_A(0) - V_B(y_B) - y_BV'_B(y_B)] \neq$
sgn$[V_A(0) - V_B(y_B)]$, even in the limit of infinitesimally
small traffic (so that $x_A + x_B$ equals some infinitesimal $\delta$),
ISPA and LB still disagree.

%This lack of ``cooperation'' among the routers in the ISPA is a major
%shortcoming of the ISPA. It reflects the fact that ISPA cannot
%successfully deal with situations where a path that ``appears''
%shorter turn out to cause congestion {\em when the actions of others
%nodes are taken into account}.  There is no mechanism built-in to the
%ISPA that will allow nodes to modify their strategies to ``cooperate''
%with the other nodes in order to improve their performance.

%As detailed above, the cost at time $t$ is $\sum_{s, d, r, p_{s, d,
%r}} x_{s, d, r, p_{s, d}}(t) [C_r(\sum_{s', d', p_{s', d'}} S_{p_{s,
%d, r}}(t))]$, where $(s, d, r, p_{s, d}, x_{s, d, r, p_{s, r, d}}(t))$
%specifies a source, a destination, a router, a path from that source to
%that destination, and a binary-valued variable giving the number of
%packets being sent through that 

\subsection{Braess' Paradox}
\label{sec:braess}
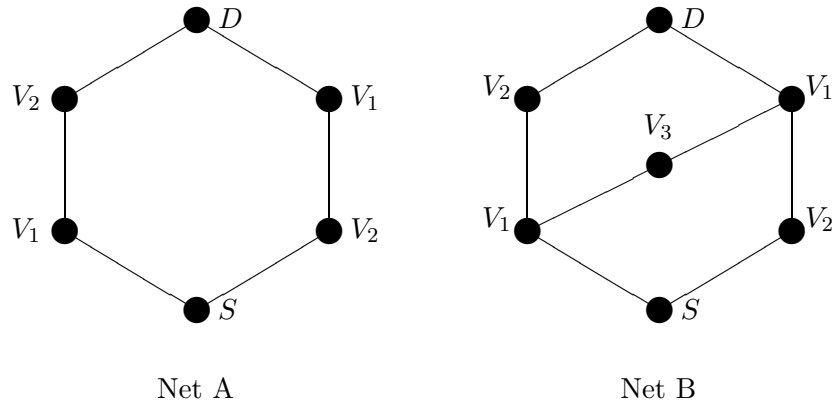
\begin{figure}[bh]
\begin{picture}(100,140)(-80,-20)
\put(50,0){\circle*{10}} 
     \put (58,0) {\makebox(-1,1)[l]{$S$}}
  \put(50,0) {\line (-5,3){50}} 
  \put(50,0) {\line (5,3){50}} 
\put(0,30){\circle*{10}}
     \put (-8,30) {\makebox(-1,1)[r]{$V_1$}}
  \put(0,30) {\line (0,1){50}} 
\put(100,30){\circle*{10}}
     \put (108,30) {\makebox(-1,1)[l]{$V_2$}}
  \put(100,30) {\line (0,1){50}} 
\put(0,80){\circle*{10}}
     \put (-8,80) {\makebox(-1,1)[r]{$V_2$}}
  \put(0,80) {\line (5,3){50}} 
\put(100,80){\circle*{10}}
     \put (108,80) {\makebox(-1,1)[l]{$V_1$}}
  \put(100,80) {\line (-5,3){50}} 
\put(50,110){\circle*{10}}
     \put (58,110) {\makebox(-1,1)[l]{$D$}}

\put(225,0){\circle*{10}}
     \put (233,0) {\makebox(-1,1)[l]{$S$}}
  \put(225,0) {\line (5,3){50}} 
  \put(225,0) {\line (-5,3){50}} 
\put(175,30){\circle*{10}}
     \put (170,30) {\makebox(-1,1)[br]{$V_1$}}
  \put(175,30) {\line (0,1){50}} 
\put(275,30){\circle*{10}}
     \put (280,30) {\makebox(-1,1)[bl]{$V_2$}}
  \put(275,30) {\line (0,1){50}} 
\put(175,80){\circle*{10}}
     \put (170,80) {\makebox(-1,1)[br]{$V_2$}}
  \put(175,80) {\line (5,3){50}} 
\put(275,80){\circle*{10}}
     \put (280,80) {\makebox(-1,1)[bl]{$V_1$}}
  \put(275,80) {\line (-5,3){50}} 
\put(225,110){\circle*{10}}
     \put (233,110) {\makebox(-1,1)[l]{$D$}}
\put(225,55){\circle*{10}}
     \put (225,65) {\makebox(-1,1)[b]{$V_3$}}
  \put(175,30){\line(2,1){100}}

\put (50,-30) {\makebox(-1,1){Net A}}
\put (225,-30) {\makebox(-1,1){Net B}}

\end{picture}

\caption{Hex network with $V_1 = 10 x \;\; ; \;\; V_2 = 50 + x \;\; 
; \;\; V_3 = 10 + x$} 
\label{fig:hex}
\end{figure}

Braess' paradox~\cite{bass92,coke90,coje97,kola98,kola99} dramatically
underscores the inefficiency of the ISPA described above.  This
apparent ``paradox'' is perhaps best illustrated through a highway
traffic example first given by Bass~\cite{bass92}: There are two
highways connecting towns S and D.  The cost associated with
traversing either highway (either in terms of tolls, or delays) is
$V_1 + V_2$, as illustrated in Net A of Figure~\ref{fig:hex}. So when
$x = 1$ (a single traveler) for either path, total accrued cost is 61
units.  If on the other hand, six travelers are split equally among
the two paths, they will each incur a cost of 83 units to get to their
destinations.  Now, suppose a new highway is built connecting the two
branches, as shown in Net B in Figure~\ref{fig:hex}. Further, note
that the cost associated with taking this highway is not particularly high
(in fact for any load higher than 1, this highway has a lower cost
than any other highway in the system).
The benefit of this highway is illustrated by the dramatically reduced
cost incurred by the single traveler: by taking the short-cut, one
traveler can traverse the network at a cost of 31 units ($2 \;V_1 + V_3$).  
Adding a new road has seemingly reduced the traversal cost dramatically.

However consider what happens when six travelers are on the highways
in net B.  If each node uses an ISPA, then at equilibrium each of the
three possible paths contains two travelers.\footnote{We have in mind
here the Nash equilibrium for this problem, where no traveler (or
equivalently, no router) can gain advantage by changing strategies.}
Due to overlaps in the paths however, this results in each traveler
incurring a cost of 92 units, which is higher than than what they
incurred {\em before} the new highway was built. The net effect of
adding a new road is to increase the cost incurred by {\em every}
traveler. This phenomenon is known as Braess' paradox.

\subsection{The Suboptimality of Load-Balancing}
\label{sec:lb}

As mentioned before, LB considers side-effects of current routing
decisions on other traffic currently being routed. However because it
does not consider side-effects of routing decisions on future traffic,
even LB may not optimize global cost averaged across all time,
depending on the details of the system. Here we present an existence
proof of this, by explicitly constructing a situation where
conventional LB is suboptimal.

Consider a
system with discrete time, in which the node $X$ under consideration must
route one packet to the (fixed) destination at each time step
(cf. Section~\ref{sec:subopt} above). Presume further that no traffic
enters any of the nodes $X$ sends to except for $X$. (So that traffic
coming from $X$ is the sole source of any costs associated with $X$'s
outbound links.) Let $S(t)$ be the number of
times our node sent a packet down some link $A$ in the $W$ time steps
preceding $t$, and take $s(t)=A,B$ to mean that the router uses link
$A$ or $B$, respectively, at time $t$. Model queue backups and the
like by having the cost to send a packet down link $A$ at time $t$ be
$C_A(S(t) / W)$, and have the cost for our router to instead send the
packet down link $B$ be $C_B(1 - S(t) / W)$,
For simplicity we assume
that both $C_A(.)$ and $C_B(.)$ are monotonically increasing functions
of their arguments.

Restrict attention to nodes that work by having $s(t) = A$ iff $S(t)
\le k$ for some real-valued threshold $k$.  The LB algorithm will
choose $s(t) = A$ iff $C_A(S(t) / W) \le C_B(1 - S(t) / W)$. So the LB
algorithm's behavior is indistinguishable from this kind of threshold
algorithm, with $k$ set so that $C_A(k / W) = C_B(1 - k/W)$. (We
implicitly assume that $C_A(.)$ and $C_B(.)$ are chosen so that such a
solution exists for $1<k<W-1$.) The question is what $k$ should
be to optimize total averaged cost across time, and in particular if
that $k$ is the same as $k_{LB}$, the $k$ that LB uses.

Now as we go from one time step to the next, the routing decision made
$W$ time steps ago drops out of the computation of $S(t)$, while the
routing decision just made is newly included. In general, $S(t+1)=
S(t) + 1$ if the router just used $A$ at time $t$ and used link $B$ at
the time $W$ time steps into the past. On the other hand, $S(t+1)=S(t)
-1$ if the router just used $B$ and used $A$ $W$ time steps ago, while
$S(t+1)=S(t)$ if the routing decision just made is the same as the
routing decision $W$ time steps ago.  So in general, $S(t)$ can only
change by -1, 0, or +1 as we go from one time step to the next.
 
Consider cases where $1 < k < W-1$, so that eventually the
router must choose an $A$, and at some subsequent time $t^*$ the
router switches from $A$ to $B$. At that time $s(t^*-1)=A$ and
$s(t^*)=B$.  This implies that $S(t^* -1) \le k, S(t^*) > k$.  Define
the value $S(t^* - 1)$ as $k^*$. Note that $S(t^*) = k^* + 1$, and $k-
1 < k^* \le k$.
 
Now for any time $t'$, if $S(t') = k^*+1$, $s(t'+1) = B$, and the only
possible next values are $S(t'+1) = k^*$ or $S(t'+1) = k^*+1$,
depending on the old decision $s(t-W)$ that gets dropped out of the
window. Similarly, if $S(t') = k^*$, $s(t'+1) = A$, and the only
possible next values are $S(t'+1) = k^*$ or $S(t'+1) = k^* + 1$,
again depending on the old decision being dropped. So we see that 
once $S(t') \in \{k^*, k^*+1\}$, it stays there forever.

This means that because of the relationship between $k$ and $k^*$,
in any interval of $W$ consecutive time steps
subsequent to $t^*$, the number of packets sent along $A$ by 
router $X$ must be $\in (k-1, k+1]$. (Note that it is possible
to send $k+1$ packets along $A$, but not $k-1$ packets.
Therefore the number sent along
$B$ must be $\in [W - (k+1), W - (k-1))$. Each time that a packet is
sent along $A$
the cost incurred is the cost of link $A$ with average traffic level
$S(t)/ W$, $C_A(S(t)/W)$. Similarly, each time the link $B$ is chosen,
the cost incurred is $C_B(1 - S(t) / W)$. Since $S(t) \in \{k^*,
k^*+1\}$, and both $C_A(.)$ and $C_B(.)$ are monotonically increasing,
the cost for sending the packet down link $A \in (C_A((k-1)/W),
C_A((k+1)/W]$, and that for sending it down link $B$ is contained in
$[C_B(1 - (k+1)/W), C_B(1 - (k-1)/W))$.

Now we know that the choice of $A$ must have average frequency (across
all time) between $k^*/W$ and $(k^*+1)/W$. Similarly, $B$ will have 
average frequency between $(1 - (k^*+1)/W)$ and $1-k^*/W$.
Accordingly, the average cost is bounded above by
\begin{eqnarray}
\frac{k^*+1}{W} C_A\left(\frac{k+1}{W}\right) \;\;+\;\;  
\left(1- \frac{k^*}{W}\right) C_B\left(1 - \frac{k-1}{W}\right) \;,
\label{eq:upper1}
\end{eqnarray}
where the first term provides the maximum possible average cost for
using link $A$, while the second term independently provides the
maximum possible average cost for using link $B$. (Note that
the actual cost will be lower since the two frequencies in this bound,
one for $A$ and one for $B$, cannot both have the values indicated.)
Because $k-1 < k^* \le k$ and since
$1-\frac{k-1}{W} = 1+\frac{2}{W}-\frac{k+1}{W}$, our upper bound
is itself bounded above by
\begin{eqnarray}
\frac{k+1}{W} C_A\left(\frac{k+1}{W}\right) \;\;+ \;\;
\left(1 + \frac{2}{W} - \frac{k+1}{W}\right) 
	C_B \left(1 + \frac{2}{W} - \frac{k+1}{W}\right) \;.
\label{eq:upper2}
\end{eqnarray}

The optimal $k$ will result in an average cost lower than the
minimum over all $k$ of the upper bound on average cost, given in 
Equation~\ref{eq:upper2}. 
So the average cost for the optimal $k$ is bounded above by the minimum 
over $k$ of this upper bound. Lable this argmin of
Equation~\ref{eq:upper2} $k$'.

Since other values of $k$ besides $k_{LB}$ result in behavior
equivalent to LB, it does not suffice to simply test if $k$' =
$k_{LB}$. Instead let us evaluate some lower bounds in a similar fashion
to how we evaluated upper bounds. Using the average 
frequencies discussed above, the average cost is bounded below by:
\begin{eqnarray}
\frac{k^*}{W} C_A\left(\frac{k-1}{W}\right) \;+\;  
\left(1-\frac{1}{W} - \frac{k^*}{W}\right) C_B\left(1 - \frac{k+1}{W}\right) \;,
\label{eq:low1}
\end{eqnarray}
where the first term provides the minimum possible average cost for
using link $A$, while the second term provides the minimum possible
average cost for using link $B$.  Again, because $k-1 < k^* \le k$, the term
is Equation~\ref{eq:low1} is further bounded below by
\begin{eqnarray}
\frac{k-1}{W} C_A\left(\frac{k-1}{W}\right) \;\;+ \;\;
\left(1-\frac{2}{W} - \frac{k-1}{W}\right) C_B\left(1 -\frac{2}{W} -
	\frac{k-1}{W}\right) \;.
\label{eq:low2}
\end{eqnarray}
In particular this bound holds for the average cost of the LB algorithm:
\begin{eqnarray}
\frac{k_{LB}-1}{W} C_A\left(\frac{k_{LB}-1}{W}\right) \;\;+ \;\;
\left(1-\frac{2}{W} - \frac{k_{LB}-1}{W}\right) 
	C_B\left(1 - \frac{2}{W} - \frac{k_{LB}-1}{W}\right) \; ,
\label{eq:lb}
\end{eqnarray}

\noindent 
where as before $k_{LB}$ satisfies $C_A(k_{LB} / W) = C_B(1 -
k_{LB}/W)$. 

By appropriate choice of $C_A(.)$ and $C_B(.)$, we can ensure that the
lower bound on the cost with the LB algorithm (Equation~\ref{eq:lb}
evaluated with $k = k_{LB}$) is higher than the upper bound on the
average cost incurred by the optimal algorithm (the minimum over $k$
of Equation~\ref{eq:upper2}).\footnote{For example, for $C_A(x) = x^2$
and $C_B(x) = x$, balancing the loads on $A$ and $B$ --- setting
$C_A(S(t)/W) = C_B(1-S(t)/W)$ --- results in $(S(t)/W)^2 = 1 -
S(t)/W$, leading to $k_{LB} / W = \frac{\sqrt{5} - 1}{2} = .618$.  For
$W = 1000$, the associated lower bound on average cost
(Equation~\ref{eq:lb}) is $.617(.617)^2 + (.998 - .617)^2 = .380$. On
the other hand, with $C_A$ and $C_B$ given as above,
Eq~\ref{eq:upper2} is $(\frac{k+1}{W})^3 \; + \; (1 + \frac{2}{W} -
\frac{k+1}{W})^2$. Differentiating with respect to $k$ and setting
the result to zero leads to $\frac{k'}{W} = -\frac{1}{3} - \frac{1}{W}
+ \frac{\sqrt{28 + 48/W}}{6}$.  For a window size of $W=1000$, this
yields $k'/W = .548$, a different result than $k_{LB}$. Plugging into
Equation~\ref{eq:upper2}, the upper bound on the performance with $k$'
is $(.549)^3 + (1.002 - .549)^2 = .371$, which is less than $.380$.}
That is, the best possible average cost achieve by load balancing will
be worse than the worst average cost that could arise through the
optimal routing strategy. This establishes that LB does not engage in
optimal routing.

\section{COIN-based Routing}
\label{sec:coin}

One common solution to these types of side-effect problems is to have
particular components of the network (e.g., a ``network manager''
\cite{kola95}) dictate certain choices to other nodes. This solution
can incur major brittleness and scaling problems however. Another kind
of approach, which avoids the problems of a centralized manager, is to
provide the nodes with extra incentives that can induce them to take
actions that are undesirable to them from a strict SPA sense. Such
incentive can be in the form of ``taxes'' or ``tolls'' added to the
costs associated with traversing particular links to discourage the
use of those links. Such schemes in which tolls are superimposed on
the nodes' goals are a special case of the more general approach of
replacing the goal of each node with a new goal. These new goals are
specifically tailored so that if they are collectively met the system
maximizes throughput. {\it A priori}, a node's goal need have no
particular relation with the SPA-type cost incurred by that node's
packets. Intuitively, in this approach, we provide each node with a
goal that is ``aligned'' with the global objective, with no separate
concern for of that goal's relation to the SPA-type cost incurred by
the traffic routed by that node.

In this section, we first summarize the theory of such systems, which
are called COllective INtelligences
(COIN's)~\cite{wowh99a,wotu99b}. We then use that theory to justify an
algorithm that only uses limited knowledge of the state of the network
(in particular knowledge that is readily available to routers in
common real data networks) to make routing decisions.  At each router,
this algorithm uses a Memory Based (MB) machine learning algorithm to
estimate the value that a private utility (provided by COIN theory)
would take on under the different candidate routing decisions. It then
makes routing decisions aimed at maximizing that utility. (We call
this algorithm an MB COIN.)

\subsection{The COIN Formalism}
\label{sec:math}

In this paper we consider systems that consist of a set of nodes,
connected in a network, evolving across a set of discrete, consecutive
time steps, $t \in \{0, 1, ...\}$.  Without loss of generality, we let
all relevant characteristics of a node $\eta$ at time $t$ ---
including its internal parameters at that time as well as its
externally visible actions --- be encapsulated by a Euclidean vector
$\underline{\zeta}_{\eta,t}$ with components
$\underline{\zeta}_{\eta,t;i}$. We call this the ``state'' of node
$\eta$ at time $t$, and let $\underline{\zeta}_{,t}$ be the state of
all nodes at time $t$, while $\underline{\zeta}$ is the state of all
node across all time.

{\bf World utility}, $G(\underline{\zeta})$, is a function of the
state of all nodes across all time. (Note that that state is a
Euclidean vector.) When $\eta$ is an agent that uses a machine
learning (ML) algorithm to ``try to increase'' its {\bf private
utility}, we write that private utility as
$g_{\eta}({\zeta})$, or more generally, to allow that
utility to vary in time, $g_{\eta,\tau}(\underline{\zeta})$.

We assume that $\underline{\zeta}$ encompasses all physically relevant
variables, so that the dynamics of the system is deterministic (though
of course imprecisely known to anyone trying to control the
system). Note that this means that $all$ characteristics of an agent
$\eta$ at $ t = 0$ that affects the ensuing dynamics of the system
must be included in $\underline{\zeta}_{\eta,0}$. For ML-based agents,
this includes in particular the algorithmic specification of its
private utility, typically in the physical form of some computer
code.  (As elaborated in \cite{wotu99b} the mathematics can be generalized
beyond ML-based agents.)

Here we focus on the case where our goal, as COIN designers, is
to maximize world utility through the proper selection of private
utility functions. Intuitively, the idea is to choose private
utilities that are aligned with the world utility, and that also have
the property that it is relatively easy for us to configure each node
so that the associated private utility achieves a large value. 
In this paper, we restrict attention to utilities of the form
$\sum_{t \ge \tau} R_{t}(\underline{\zeta}_{,t})$ for {\bf reward
functions} $R_t$ (simply $\sum_{t} R_{t}(\underline{\zeta}_{,t})$ for
non-time-varying utilities). From now on, we will only consider
world utilities whose associated set of \{$R_t$\} are all time-translations of
one another. In particular, as shown below, overall network throughput
is expressible this way.

We need a formal definition of the concept of having private
utilities be ``aligned'' with $G$. Constructing such a formalization
is a subtle exercise. For example, consider systems where the world
utility is the sum of the private utilities of the individual
nodes. This might seem a reasonable candidate for an example of
``aligned'' utilities. However such systems are examples of the more
general class of systems that are ``weakly trivial''. It is well-known
that in weakly trivial systems each individual agent greedily trying
to maximize its own utility can lead to the tragedy of the commons
~\cite{hard68,crow69} and actually {\it minimize} $G$.  In particular, this can be
the case when private utilities are independent of time and $G =
\sum_{\eta} g_{\eta}$. Evidently, at a minimum, having $G =
\sum_{\eta} g_{\eta}$ is not sufficient to ensure that we have
``aligned'' utilities; some alternative formalization of the concept
is needed.\footnote{Note that in the simple network discussed in
Section~\ref{sec:ispa}, the utilities are weakly trivial, since
$G(\vec{x}, \vec{y}) = g_X(\vec{x}) + g_y(\vec{y})$. This provides
another perspective on the suboptimality of ISPA in that network.}

A more careful alternative formalization of the notion of aligned
utilities is the concept of ``factored'' systems.  A system is {\bf
factored} at time $\tau$ when the following holds for each agent
$\eta$ individually: A change at time $\tau$ to the state of $\eta$ alone,
when propagated across time, will result in an increased value of
$g_{\eta,\tau}(\underline{\zeta})$ if and only if it results in an increase
for $G(\underline{\zeta})$~\cite{wotu99b}.

For a factored system, the side-effects of a change to $\eta$'s $t =
\tau$ state that increases its private utility cannot decrease world
utility. There are no restrictions though on the effects of that
change on the private utilities of other nodes and/or times. In
particular, we don't preclude different a node's algorithm at two
different times from ``working at cross-purposes'' to each other, so
long as at both moments the node is working to improve $G$.  In
game-theoretic terms, optimal global behavior corresponds to the
agents' reaching a private utility Nash equilibrium for such
systems~\cite{futi91}.  In this sense, there can be no TOC for a
factored system. As a trivial example, a system is factored for
$g_{\eta,\tau} = G \; \forall \eta$.

Define the {\bf effect set} of the node-time pair $(\eta,\tau)$ at
$\underline{\zeta}$, $C^{eff}_{(\eta,\tau)}(\underline{\zeta})$, as the set of
all components $\underline{\zeta}_{\eta',t}$ which under the forward
dynamics of the system have non-zero partial derivative with respect to the
state of node $\eta$ at $t=\tau$.  Intuitively, $(\eta,\tau)$'s effect
set is the set of all components $\underline{\zeta}_{\eta',t \ge \tau}$ which
would be affected by a change in the state of node $\eta$ at time
$\tau$. (They may or may not be affected by changes in the $t=\tau$
states of the other nodes.)

Next, for any set $\sigma$ of components ($\eta', t$), define
$\mbox{CL}_\sigma(\underline{\zeta})$ as the ``virtual'' vector formed
by clamping the $\sigma$-components of $\underline{\zeta}$ to an
arbitrary fixed value. (In this paper, we take that fixed value to be
$\vec{0}$ for all components listed in $\sigma$.) 
The value of the effect set {\bf wonderful life utility} (WLU for short) 
for $\sigma$ is defined as:
\begin{equation}
WLU_{\sigma}(\underline{\zeta}) \equiv 
G(\underline{\zeta}) - G(\mbox{CL}_{\sigma}(\underline{\zeta}))
.
\end{equation}
\noindent
In particular, we are interested in the WLU for the effect set of
node-time pair $(\eta,\tau)$. This WLU is the difference between the
actual world utility and the virtual world utility where all node-time
pairs that are affected by $(\eta,\tau)$ have been clamped to a zero
state while the rest of $\underline{\zeta}$ is left unchanged.

Since we are clamping to $\vec{0}$, we can view $(\eta,\tau)$'s effect set
WLU as analogous to the change in world utility that would have arisen
if $(\eta,\tau)$ ``had never existed''.  (Hence the name of this utility
- cf. the Frank Capra movie.)  Note however, that $\mbox{CL}$ is a
purely ``fictional'', counter-factual operator, in the sense that it
produces a new $\underline{\zeta}$ without taking into account the
system's dynamics.  The sequence of states the node-time pairs in
$\sigma$ are clamped to in constructing the WLU need not be consistent
with the dynamical laws of the system.  This dynamics-independence is
a crucial strength of the WLU.  It means that to evaluate the WLU we
do {\it not} try to infer how the system would have evolved if node
$\eta$'s state were set to $\vec{0}$ at time $\tau$ and the system evolved from
there. So long as we know $\underline{\zeta}$ extending over all time,
and so long as we know $G$, we know the value of WLU.

Assuming our system is factored with respect to private utilities
\{$g_{\eta,\tau}$\}, we want each node to be in a state at time $\tau$
that induces as high a value of the associated private utility as
possible (given the initial states of the other nodes). Assume $\eta$
is ML-based and able to achieve fairly large values of most private
utilities we are likely to set it for time $\tau$, i.e., assume that
given that private utility $g_{\eta,\tau}$, the rest of the
components of $\underline{\zeta}_{\eta,\tau}$ are set by $\eta$'s
algorithm in such a way so as to achieve a relatively high value of
$g_{\eta,\tau}$. 
So our problem becomes
determining for what \{$g_{\eta,\tau}$\} the nodes will best be
able to achieve high $g_{\eta}$ (subject to each other's actions)
while also causing dynamics that is factored for $G$ and the
\{$g_{\eta,\tau}$\}.

As mentioned above, regardless of the system dynamics, having
$g_{\eta,\tau} = G \; \forall \eta$ means the system is factored at
time $\tau$. It is also true that regardless of the dynamics,
$g_{\eta,\tau} = WLU_{C^{eff}_{(\eta,\tau)}} \; \forall \eta$ is a
factored system at time $\tau$ (proof in ~\cite{wotu99b}). Which of 
these two choices of the $\{g_{\eta,\tau}\}$ should we use?

To answer this, note that since each agent is operating in a large
system, it may experience difficulty discerning the effects of its
actions on $G$ when $G$ sensitively depends on all the myriad
components of the system. Therefore each $\eta$ may have difficulty
learning from past experience what to do to achieve high
$g_{\eta,\tau}$ when $g_{\eta,\tau} = G$.\footnote {In
particular, in the routing problem, having private rewards given by
the world reward functions means that to provide each router with its
reward at each time step we need to provide it the full throughput of
the entire network at that step. This is usually infeasible in
practice. Even if it weren't though, using these private utilities
would mean that the routers face a very difficult task in trying to
discern the effect of their actions on their rewards, and
therefore would likely be unable to learn their best routing
strategies.}

This problem can be mitigated by using effect set WLU as the private
utility, since the subtraction of the clamped term removes much of the
``noise'' of the activity of other agents, leaving only the underlying
``signal'' of how the agent in question affects the utility.  (This
reasoning is formalized as the concept of ``learnability'' in
~\cite{wotu99b}.)  Accordingly, one would expect that setting private
utilities to WLU's ought to result in better performance than having
$g_{\eta,\tau} = G \; \forall \eta,\tau$. In practice, we will
often only be able to estimate the ``primary'', most prominent portion
of the effect set. However assuming that the associated WLU is close
enough to being factored, we would expect the advantage in
learnability with such a WLU to still result in better performance
than would using $g_{\eta,\tau} = G \; \forall \eta,\tau$. (See
~\cite{wowh99a,wotu99b}.) Indeed, for the sake of improving
learnability, often we will elect to exclude certain node-time pairs
from our estimate of the effect set of $(\eta,\tau)$, even if we are
sure that that are affected by $\underline{\zeta}_{\eta,\tau}$. This
will be the case if we expect that the changes in $G$ due to varying
$\underline{\zeta}_{\eta,\tau}$ that are ``mediated'' through those
node-time pairs are relatively insignificant, and therefore
effectively constitute noise for the learning process, so that their
effect on learnability is more important than their effect on factoredness.

\subsection{Model Description} \label{sec:model} 

To apply the COIN formalism to a network routing model, we must
formally identify the components of that model as deterministically
evolving vectors $\underline{\zeta}_{,t}$. In the model used in this
paper, at any time step all traffic at a router is a set of pairs of
integer-valued traffic amounts and associated ultimate destination
tags.  At each such time step $t$, each router $r$ sums the
integer-valued components of its current traffic at that time step to
get its {\bf instantaneous load}. We write that load as $z_r(t) \equiv
\sum_d x_{r,d}(t)$, where the index $d$ runs over ultimate
destinations, and $x_{r,d}(t)$ is the total traffic at time $t$ going
from $r$ towards $d$.  After its instantaneous load at time $t$ is
evaluated, the router sends all its traffic to the next downstream
routers, in a manner governed by its routing algorithm.  We indicate
such ``next routers'' by writing $x_{r,d}(t) = \sum_{r'}
x_{r,d,r'}(t)$ where $r'$ is the first stop on the path to be followed
from router $r$ to ultimate destination $d$.  After all such routed
traffic goes to those next downstream routers, the cycle repeats itself,
until all traffic reaches its destinations.

In our simulations, for simplicity, traffic was only introduced into
the system (at the {\bf source routers}) at the beginning of
successive {\bf waves} of $L$ consecutive time steps. ($L$ was always
chosen to be the minimal number necessary for all traffic to reach its
destination before the next wave of traffic is initiated.) We use
$\kappa(t)$ to indicate either the integer-valued wave number
associated with time $t$ or the set of all times in that wave, as the
context indicates.

In a real network, the cost of traversing a router does not change
dramatically from one packet to the next. To simulate this effect, we
use time-averaged values of the load at a router rather than
instantaneous load to determine the cost a packet incurs in traversing
that router.  More formally, we define the router's {\bf windowed
load}, $Z_r(t)$, as the running average of that router's load value
over a window of the previous $W$ timesteps: $Z_r(t) \equiv
\frac{1}{W} \sum_{t'=t - W + 1}^{t} z_r(t') = \sum_{d'} X_{r,d'}(t)$,
where the value of $X_{r,d}(t)$ is set by the dynamical law
$X_{r,d}(t) = \frac{1}{W} \sum_{t' = t - W + 1}^t x_{r,d}(t'))$. ($W$
is always set to an integer multiple of $L$.) For large enough $W$,
using such a window means that in a typical scenario the costs across
nodes will only change substantially over time scales significantly
larger than that of the individual routing decisions.  The windowed
load is the argument to a {\bf load-to-cost} function, $V(\cdot)$,
which provides the {\bf cost} accrued at time $t$ by each packet
traversing the router at this timestep.  That is, at time $t$, the
cost for each packet to traverse router $r$ is given by
$V(Z_r(t))$.\footnote{Note that in our model, the costs are accrued at
the routers, not the links. Also note that for simplicity we do not
physically instantiate the cost as a temporal delay in crossing a
router.}  (We also introduce ``dummy nodes'' denoted by $V_0(\cdot) = 0$ 
which help in translating the mathematics into the simulations. Omitting
them will have no effect on the simulations.)
Different routers have different $V(\cdot)$, to reflect the
fact that real networks have differences in router software and
hardware (response time, queue length, processing speed etc). For
simplicity, $W$ is the same for all routers however. With these
definitions, world utility is given by
\begin{eqnarray} 
G(\underline{\zeta}) =& \sum_{t,r} \; \; z_r(t) \; \; V_r(Z_r(t)) \nonumber \\ \nonumber 
=& \sum_{t,r,d} x_{r,d}(t) V_r(Z_r(t)) \\ \nonumber
 =& \sum_{t,r,d} x_{r,d}(t) V_r(\sum_{d'} X_{r,d'}(t)) \;.  
\end{eqnarray}

Our equation for $G$ explicitly demonstrates that, as claimed above, in our
representation we can express $G(\underline{\zeta})$ as a sum of
rewards, $\sum_t R_t(\underline{\zeta}_{,t})$, where
$R(\underline{\zeta}_{,t})$ can be written as function of a pair of
$(r,d)$-indexed vectors: $R_t(x_{r,d}(t), X_{r,d}(t)) = \sum_{r,d}
x_{r,d}(t) V_r(\sum_{d'} X_{r,d'}(t))$. Also as claimed, the $R_t$ are
temporal translations of one another.

Given this model, some of the components of $\underline{\zeta}_{,t}$
must be identified with the values $x_{r,d,r'}(t) \; \forall \; r, d,
r'$ and $t$, since those are the actions we will take.  Since all
arguments of $G$ must be components of $\underline{\zeta}$, we also
include the $X_{r,d}(t) \; \forall r,d,t$ as components of
$\underline{\zeta}_{,t}$. (We could use the \{$Z_r(t)$\} as an
alternative, but this would provide a ``coarser'' WLU; see below.)
Formally, for routing based on ML agents, other variables must also be
included in $\underline{\zeta}$, to capture the (deterministically
evolving) internal parameters used by those agents to make their
routing decisions. We won't have any need to explicitly delineate such
variables here however, and will mostly phrase the discussion as
though there were no such internal parameters.

Now the values \{$x_{r,d,r'}(t-1)\} \; \forall r,d,r'$ specify the
values \{$x_{r,d}(t)\} \; \forall r, d$ directly. However the
decisions of each router's algorithm at all times $t$ is a fixed
function of the \{$x_{r,d}(t-1)$\} and the \{$Z_{r}(t-1) = \sum_{d'}
X_{r,d'}(t-1)$\}, a function given by the routing algorithm which is
implicitly encapsulated in the dynamical laws governing the system. So
in point of fact we can map the set of \{$x_{r,d,r'}(t-1)\} \; \forall
r,d,r'$ to the full set \{$x_{r,d,r'}(t)\} \; \forall r,d,r'$, not
just to \{$x_{r,d}(t)$\}. Accordingly, the $x_{r,d,r'}$ undergo
deterministic evolution. Since their values across time set all the
values of the $X_{r,d}(t)$ across time, we see that the entire set of
the components of $\underline{\zeta}_{,t}$ undergo deterministic
evolution in this representation, as required.

For evaluating the wonderful life utility we will need to group the
components of $\underline{\zeta}_{,t}$ into disjoint nodes
$\eta$. Here we will have two types of node, both types being indexed
by router-destination pairs. For each such node index $(r, d)$, the
first node type is the variable $X_{r,d}(t)$, and the second node type
is the Euclidean vector with components indexed by $r'$,
$(x_{r,d})_{r'}(t)$. In setting ``actions'' we are concerned with
setting the states of the nodes of the second type. Accordingly, our
learners will all be associated with nodes of this second type. Unless
explicitly indicated otherwise, from now on we will implicitly have
that second type of node in mind whenever we refer to a ``node'' or
use the symbol $\eta$.

At time step $t$, ISPA has access to all the windowed loads at time
step $t - 1$ (i.e., it has access to $Z_{r}(t-1) \; \forall r$), and
assumes that those values will remain the same at all times $\ge
t$. (Note that for large window sizes and times close to $t$, this
assumption is arbitrarily accurate.)  Using this assumption, in
ISPA, each router sends packets along the path that it calculates will
minimize the costs accumulated by its packets.

\subsection{COIN Routing}
\label{sec:coinroute}
Based on the COIN formalism presented in Section~\ref{sec:math} and the model
described above, we now present the COIN-based routing algorithms.

To evaluate the WLU for a node $(r,d)$ at any time $\tau$, we must
estimate the (primary members of the) associated effect set. This
means determining what components of $\underline{\zeta}_{,}$ will,
under the dynamics of the system, be changed by altering any of the
components of the vector $x_{r,d}(\tau)$. 

As a first approximation, we will ignore effects that changing
$x_{r,d}(\tau)$ may have that are ``mediated'' by the learning
algorithms running in the system. That is, we ignore changes that
arise due to the the effects of changing $x_{r,d}(\tau)$ on rewards,
changes which induce changes in future training sets, which then in
turn get mapped to changes in the \{$x_{r,d,r'}(t)$\} (and therefore
the \{$X_{r,d}(t)$\}) via the learning algorithms running on the
nodes.

As another approximation, we will ignore effects mediated by the
routing algorithms' observations of the state of the network. That is,
we ignore changes in the \{$x_{r'',d',r'''}(t)$\} that varying
$x_{r,d}(\tau)$ may cause due to $(r'',d')$'s routing algorithm
perceiving a different state of the network and modifying its routing
decisions accordingly. We only consider the behavior of those routing
algorithms that are (potentially) directly affected by $x_{r,d}(\tau)$
in that they (potentially) have to route packets that, at time $\tau$,
passed through $r$ on the way to $d$.  So in particular we ignore
effects of $x_{r,d}(\tau)$ on the \{$x_{r'',d' \neq d,r'''}(t)$.

Since all packets routed in a wave arrive at their destinations by the
end of the wave, these approximations mean that the only
$x_{r'',d'',r'''}(t)$ that are in our estimate for $x_{r,d}(\tau)$'s
effect set have $t$ in the same wave as $\tau$. (These are the only
ones that are, potentially, directly affected by the
\{$x_{r,d,r'}(t)$\} by ``chaining together'' the sequence of
$x_{r'',d'',r'''}(t)$ that get the packets in $x_{r,d}(t)$ to their
ultimate destination.)  Due to the wave nature of our simulations
though, the only $x_{r'',d'',r'''}(t)$ within $\tau$'s wave that are
affected by $x_{r,d}(\tau)$ all have $d'' = d$. For reasons of coding
simplicity, we do not concern ourselves whether $t < \tau$ within a
given wave and then exclude some $x_{r'',d'',r'''}(t)$ accordingly. In
other words, all $t$ within $\tau$'s wave are treated equally.

So one set of members of $x_{r,d}(\tau)$'s effect set is
\{$x_{r'',d,r'''}(t) \; \forall r'',d,r''',t \in \kappa(\tau)$\}. Note
that some of these members will be relatively unaffected by
$x_{r,d}(\tau)$ (e.g., those with $r''$ far in the net away from
$r$). Again for simplicity, we do not try to determine these and
exclude them. As with keeping the $x_{r'',d,r'''}(t<\tau)$, this
inclusion of extra nodes in our estimate of the effect set should hurt
learnability, but in general should not hurt factoredness. Therefore
it should delay how quickly the learners determine their optimal
policies, but it won't affect how the quality (for $G$) of those
polices finally arrived at. Note also that trying to determine whether
some particular $x_{r'',d,r'''}(t \in \kappa(\tau))$ should be
included in $x_{r,d}(\tau)$'s effect set would mean, in part,
determining whether packets routed from $(r,d)$ would have reached
$r''$ if $(r,d)$ had made some routing decision different from the one
it actually made. This would be a non-trivial exercise, in general.

In contrast to the case with the $x_{r'',d',r'''}(t)$, there are
$X_{r'',d'}(t)$ with $t$ in the future of $\tau$'s wave that both are
affected by $x_{r,d}(t)$ and also are not excluded by any of our
approximations so far. In particular, the $X_{r'',d}(t)$ with either
$r'' = r$ or $r''$ one hop away from $r$ will be directly affected by
$x_{r,d}(t)$, for $t \in \cup_{i=0}^{W-1} \kappa(\tau + iL))$ (cf. the
definition of the $X$ variables). For simplicity, we restrict
consideration of such $X_{r'',d}$ variables to those with the same
router as $r$, $r'' = r$.

This final estimate for the effect set is clearly rather poor ---
presumably results better than those presented below would accrue to
use of a more accurate effect set. However it's worth bearing in mind
that there is a ``self-stabilizing'' nature to the choice of effect
sets, when used in conjunction with effect set WLU's.  This nature is
mediated by the learning algorithms. If I take two nodes and give them
the same utility function, then the reward one node gets will be
determined in part by what the other one does. So as it modifies its
behavior to try to increase its reward, that first node will be
modifying its behavior in a way dependent on what the other node
does. In other words, if two nodes are given the same WLU because they
are estimated to be in each other's effect set, then {\it ipso facto}
they will be in each other's effect set.

Using our estimate for the effect
set, the WLU for $(\eta,\tau)$ is given by the difference between the
total cost accrued in $\tau$'s wave by all nodes in the network and
the cost accrued by nodes when all nodes sharing $\eta$'s destination
are ``erased.'' More precisely, any node $\eta$ that has a destination
$d$ will have the following effect set WLU's, $g_{\eta,\tau}$:
\begin{eqnarray}
g_{\eta,\tau}(\underline{\zeta}) \!  \! \! \!  \! &=& G(\underline{\zeta}) -
  G(\mbox{CL}_{C^{eff}_{(\eta,\tau)}}(\underline{\zeta})) \label{eq:geta} \nonumber \\ 
& = & \! \! 
\sum_{t,r',d'} \; x_{r',d'}(t) \;  V_{r'}\left(\sum_{d'} X_{r',d'}(t)\right)
\; - \sum_{t, r',d'} \; \left[ x_{r',d'}(t) (1 - I(t \in
       \kappa(\tau))I(d' = d)) \right]    \nonumber \\
&& \; \times \;\;\;\;\;\; V_{r'} \left( \sum_{d''} \;\;
      [\;X_{r',d''}(t)\;\;(1 - I(t \in \cup_{i=0}^{W-1}
      \kappa(\tau + iL)) I(d'' = d))\;] \right) \nonumber \\ 
& = &  \! \! \sum_{t \in \kappa(\tau)} \sum_{r'} \; \left( \sum_{d'} \; x_{r',d'}(t) \; \;
 V_{r'}(\sum_{d''} X_{r',d''}(t)) \; - \; \sum_{d' \neq d}  x_{r',d'}(t)  \;
 V_{r'}(\sum_{d'' \neq d} X_{r',d''}(t)) \right)  \nonumber \\
&& + \! \! \sum_{t \in \cup_{i=1}^{W-1} \kappa(\tau + iL)}
\sum_{r'} \left( \sum_{d'} x_{r',d'}(t) \;[V_{r'}(\sum_{d''} X_{r',d''}(t)) -
V_{r'}(\sum_{d'' \neq d} X_{r',d''}(t))] \right) 
\label{eq:netwlu}
\end{eqnarray} 
\noindent
where $I(.)$ is the indicator function that equals 1 if its argument
is true, 0 otherwise.

To allow the learner to receive feedback concerning its actions in a
wave immediately following that wave rather than wait for $\sim WL$
time steps, we will approximate the second sum in that last equality,
the one over times following $\tau$'s wave, as zero. There is another
way we can view the resultant expression, rather than as an
approximation to the effect set WLU. That is to view it as the exact
WLU of an approximation to the effect set, an approximation which
ignores effects on future windowed loads of clamping a current traffic
level. Regardless of what view we adopt, presumably better performance
could be achieved if we did not implement this approximation. 

Given this approximation, our WLU becomes a wave-indexed
time-translation-invariant WL ``reward function'' (WLR):
\begin{eqnarray}
g_{\eta,\tau}(\underline{\zeta}_{,t \in \kappa(\tau)})  = 
\! \!   \sum_{t \in \kappa(\tau), r'}  \! 
\left(   \sum_{d'} \; x_{r',d'}(t) 
\; V_{r'}(\sum_{d''} X_{r',d''}(t)) \right. \nonumber \\ 
\! \! \! \! \! \! \! \! \!  -    \left.  \sum_{d' \neq d} \; 
x_{r',d'}(t) \; \; V_{r'}(\sum_{d'' \neq d} X_{r',d''}(t)) \right)  .
\end{eqnarray}

\noindent
Notice that traffic going
from a router $r' \neq r$ to a destination $d' \neq d$ affects the
value of the WLR for node $(r,d)$. This reflects the fact that WLR
takes into account side-effects of $(r,d)$'s actions on other nodes.
Note also that each $r'$-indexed term contributing to the WLR can be
computed by the associated router $r'$ separately, from information
available to that router.  Subsequently those terms can be propagated
through the network to $\eta$, in much the same way as routing tables
updates are propagated.

Given this choice of private utility,
we must next specify how the COIN-based routing algorithm collects the
initial data that (in conjunction with this utility) is to be used to
guide the initial routing decisions that every node with more than one routing
option must make. In our experiments that data was
collected during a preliminary running of an ISPA.  In this
preliminary stage, the routing decisions are made using the ISPA, but
the resulting actions are ``scored'' using the WLR given by
Equation~\ref{eq:netwlu}.  \footnote{We use the ISPA to generate the
routing decisions in the initial data since it is likely in practice
that some kind of SPA will be the routing algorithm running prior to
``turning on'' the COIN algorithm. Alternately one can generate the
initial data's routing decisions by having the routers make random
decisions, or by having them implement a sequence of decisions that
``sweeps'' across a grid through the possible set of actions.}  The
data collected in this stage provides us with initial input-output
training sets to be used by the machine learning algorithm on each
node: for each router-destination node, inputs are identified with
windowed loads on outgoing links, and the associated WLR values for
the destination in question are the outputs.

After sufficient initial data is collected using the ISPA, the system
switches to using the COIN algorithm to make subsequent routing
decisions.  In this stage, each node routes packets along the link
that it estimates (based on the training set) would provide the best
WLR.  To perform the estimation, the MB COIN makes use of a
single-nearest-neighbor algorithm as its learner. This algorithm
simply guesses that the output that would ensue from any candidate
input is the same as the output of the element of the training set
that is the nearest neighbor (in input space) of that candidate
input.\footnote{This is a very simple learning algorithm, and we use
it here only to demonstrate the potential practical feasibility of a
COIN-based routing algorithm. The performance can presumably be improved if more
sophisticated learning algorithms (e.g., Q-learning
\cite{suba98,wada92}) are used.}  In other words, the learner finds
the training set input-output pair whose input value (loads on
outgoing links) is closest to that which would result from each
potential routing decision. Then the learner assigns the WLR
associated with that training data pair as the estimate for what WLR
would result from said routing decision. These WLR values are then
used to choose among those potential routing decisions. The
input-output data generated under this algorithm is adding to the
training set as it is generated.

In this routing algorithm, the routers only estimate how their routing
decisions (as reflected in their loads at individual time steps) will
affect their WLR values (based on many nodes' loads).  It is also
possible to calculate {\it exactly} how the routing decisions affect
the routers' WLR's if, unlike the MB COIN, we had full knowledge of
the loads of all nodes in the system. In a way similar to ISPA, for
each router we can evaluate the exact WLR value that would ensue from
each of its candidate actions, under the assumption that windowed
loads on all other routers are the same one wave into the future as
they are now. We call this algorithm for directly maximizing WLR (an
algorithm we call the full knowledge COIN (FK COIN)).

Note that under the assumption behind the FK COIN, the action $\eta$
chooses in wave $\kappa(\tau)$ that maximizes WLR will also maximize
the world reward. In other words, WL reward is perfectly factored with
respect to (wave-indexed) world reward, even though the associated
utilities are not related that way (due to inaccuracy in our estimate
of the effect set).  Due to this factoredness, the FK COIN is
equivalent to load balancing on world rewards. Since LB in general
results in inferior performance compared to LB over time, and since
the FK COIN is equivalent to LB, one might expect that its performance
is suboptimal. Intuitively, this suboptimality reflects the fact that
one should not choose the action only with regard to its effect on
current reward, but also with concern for the reward of future
waves. In the language of the COIN framework, this suboptimality can
be viewed as a restatement of the fact that for our inexactly
estimated effect set, the system will not be perfectly factored.

The learning algorithm of the MB COIN as described is extraordinarily
crude. In addition, the associated scheme for choosing an action is
purely exploitative, with no exploration whatsoever. Rather than
choose some particular more sophisticated scheme and tune it to fit
our simulations, we emulated using more sophisticated algorithms {\it
in general}. We did this by modifying the MB COIN algorithm to
occasionally have the FK COIN determine a router's action rather than
the purely greedy learner outlined above. The {\bf steering parameter}
discussed in Section~\ref{sec:steer} determines how often the routing
decision is based on the MB COIN as opposed to the FK COIN.

\section{SIMULATION RESULTS}
\label{sec:sim} 
Based on the model and routing algorithms discussed above, we have
performed simulations to compare the performance of ISPA and MB COIN
across a variety of networks, varying in size from five to eighteen
nodes. In all cases traffic was inserted into the network in a
regular, non-stochastic manner at the sources.  The results we report
are averaged over 20 runs. We do not report error bars as they are
all lower than $0.05$.

In Sections~\ref{sec:bootes}~-~\ref{sec:ray} we analyze traffic
patterns over four networks where ISPA suffers from the Braess'
paradox. In contrast, the MB COIN almost never falls prey to the paradox 
for those
networks (or for no networks we have investigated is the MB COIN 
	     significantly susceptible to Braess' paradox).
Then in Section~\ref{sec:steer} we discuss
the effect on the MB COIN's performance of the ``steering'' parameter
which determines the intelligence of the MB COIN.\footnote{In
Sections~\ref{sec:bootes}~-~\ref{sec:ray}, the steering parameter is
set at 0.5.}

\subsection{Bootes Network}
\label{sec:bootes} 
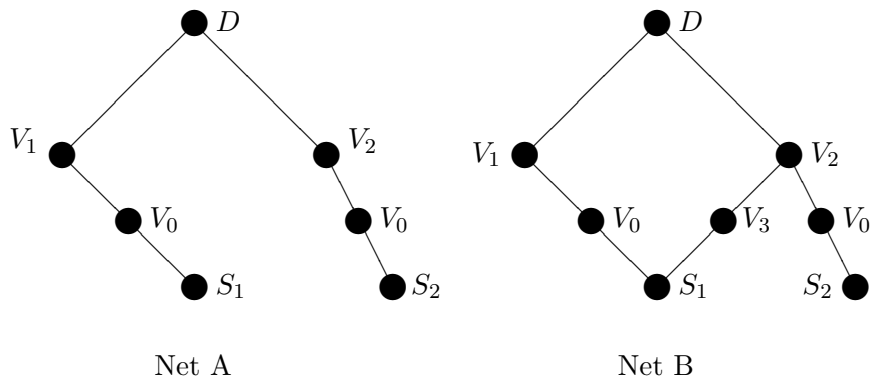
\begin{figure} [htb]
\begin{picture}(100,140)(-70,-30)
\put(50,0){\circle*{10}} 
     \put (58,0) {\makebox(-1,1)[l]{$S_1$}}
  \put(50,0) {\line (-1,1){50}} 
\put(0,50){\circle*{10}}
     \put (-8,55) {\makebox(-1,1)[r]{$V_1$}}
\put(100,50){\circle*{10}}
     \put (108,55) {\makebox(-1,1)[l]{$V_2$}}
  \put(0,50) {\line (1,1){50}} 
  \put(100,50) {\line (-1,1){50}} 
\put(50,100){\circle*{10}}
     \put (58,100) {\makebox(-1,1)[l]{$D$}}
\put(125,0){\circle*{10}}
     \put (132,0) {\makebox(-1,1)[l]{$S_2$}}
  \put(125,0) {\line (-1,2){25}}

\put(25,25){\circle*{10}} 
     \put (33,25) {\makebox(-1,1)[l]{$V_0$}}
\put(112,25){\circle*{10}} 
     \put (120,25) {\makebox(-1,1)[l]{$V_0$}}

\put(225,0){\circle*{10}} 
     \put (233,0) {\makebox(-1,1)[l]{$S_1$}}
  \put(225,0) {\line (-1,1){50}} 
  \put(225,0) {\line (1,1){50}} 
\put(250,25){\circle*{10}}
     \put (257,25) {\makebox(-1,1)[l]{$V_3$}}
\put(175,50){\circle*{10}}
     \put (167,50) {\makebox(-1,1)[r]{$V_1$}}
\put(275,50){\circle*{10}}
     \put (283,50) {\makebox(-1,1)[l]{$V_2$}}
  \put(175,50) {\line (1,1){50}} 
  \put(275,50) {\line (-1,1){50}} 
\put(225,100){\circle*{10}}
     \put (233,100) {\makebox(-1,1)[l]{$D$}}
\put(300,0){\circle*{10}}
     \put (280,0) {\makebox(-1,1)[l]{$S_2$}}
  \put(300,0) {\line (-1,2){25}}

\put(200,25){\circle*{10}} 
     \put (208,25) {\makebox(-1,1)[l]{$V_0$}}
\put(287,25){\circle*{10}} 
     \put (295,25) {\makebox(-1,1)[l]{$V_0$}}

\put (50,-30) {\makebox(-1,1){Net A}}
\put (225,-30) {\makebox(-1,1){Net B}}

\end{picture}
\caption{Bootes Network}
\label{fig:bootes}
\end{figure}

The first network type we investigate is shown in Figure~\ref{fig:bootes}.
It is in many senses a trivial network. (Indeed, in Net A, 
the sources do not even have any choices to make.)  
The loads introduced at the sources do not change in time and are listed
in Tables~\ref{tab:bootes2} and \ref{tab:bootes4}, along with the
performances of our algorithms.

\begin{table}[htb]  \centering
\caption{Average Per Packet Cost for BOOTES2 networks 
for $V_1 = 10 + log(1 + x) \; ; \; V_2 =  4 x^2 \; ; \; V_3 = log(1 + x) $ .}
\vspace*{.1in}
\begin{tabular}{c|c|c|c} \hline
Loads at $(S_1,S_2)$ & Net &  ISPA & MB COIN \\ \hline \hline
1,1 & A &         6.35 &      6.35 \\
    & B &         8.35 &      5.93 \\ \hline
2,1 & A &         8.07 &      8.07 \\
    & B &        10.40 &      7.88 \\ \hline
2,2 & A &         9.55 &      9.55 \\
    & B &        10.88 &      9.71 \\ \hline
4,2 & A &        10.41 &     10.41 \\ 
    & B &        11.55 &     10.41 \\ \hline
\end{tabular}
\label{tab:bootes2}
\end{table}

\begin{table}[htb]  \centering
\caption{Average Per Packet Cost for BOOTES4 network 
for $V_1 = 50 + log(1 + x) \; ; \; V_2 =  10 x \; ; \; V_3 = log(1 + x) $ .}
\vspace*{.1in}
\begin{tabular}{c|c|c|c} \hline
Loads at $(S_1,S_2)$ & Net &  ISPA & MB COIN \\ \hline \hline
1,1 & A &        30.35 &      30.35 \\
    & B &        20.35 &      20.35 \\ \hline
2,2 & A &        35.55 &      35.55 \\
    & B &        40.55 &      34.99 \\ \hline
4,2 & A &        41.07 &      41.07 \\
    & B &        50.47 &      44.13 \\ \hline
6,3 & A &        44.63 &      44.63 \\ 
    & B &        51.40 &      44.63 \\ \hline
\end{tabular}
\label{tab:bootes4}
\end{table}

The MB COIN results are identical to the ISPA results
in the absence of the additional link (Network A). 
However,  Braess' paradox arises with ISPA, in that the addition
of the new link in network B degrades the performance of the ISPA
in six of the eight traffic regimes and load-to-cost functions investigated.
The MB COIN on the other hand is only hurt by the addition of the new link
once, and manages to gainfully exploit it seven times. 
(When behavior is analyzed infinitesimally, the MB COIN either uses the
  additional link efficiently or chooses to ignore it in those seven
  cases.) Moreover, the MB COIN's performance with the additional link is
always better than the ISPA's.
For example, adding the new link causes a degradation of the performance by
as much as 30 \% (loads = $\{2,1\}$) for the ISPA, whereas for the same load
vector MB COIN performance improves by 7 \%.

\subsection{Hex Network}
\begin{figure}[bth]
\begin{picture}(100,140)(-80,-20)
\put(50,0){\circle*{10}} 
     \put (58,0) {\makebox(-1,1)[l]{$S$}}
  \put(50,0) {\line (-5,3){50}} 
  \put(50,0) {\line (5,3){50}} 
\put(0,30){\circle*{10}}
     \put (-8,30) {\makebox(-1,1)[r]{$V_1$}}
  \put(0,30) {\line (0,1){50}} 
\put(100,30){\circle*{10}}
     \put (108,30) {\makebox(-1,1)[l]{$V_2$}}
  \put(100,30) {\line (0,1){50}} 
\put(0,80){\circle*{10}}
     \put (-8,80) {\makebox(-1,1)[r]{$V_2$}}
  \put(0,80) {\line (5,3){50}} 
\put(100,80){\circle*{10}}
     \put (108,80) {\makebox(-1,1)[l]{$V_1$}}
  \put(100,80) {\line (-5,3){50}} 
\put(50,110){\circle*{10}}
     \put (58,110) {\makebox(-1,1)[l]{$D$}}

\put(0,55){\circle*{10}}
     \put (-8,55) {\makebox(-1,1)[r]{$V_0$}}
\put(100,55){\circle*{10}}
     \put (106,55) {\makebox(-1,1)[l]{$V_0$}}

\put(225,0){\circle*{10}}
     \put (233,0) {\makebox(-1,1)[l]{$S$}}
  \put(225,0) {\line (5,3){50}} 
  \put(225,0) {\line (-5,3){50}} 
\put(175,30){\circle*{10}}
     \put (170,30) {\makebox(-1,1)[br]{$V_1$}}
  \put(175,30) {\line (0,1){50}} 
\put(275,30){\circle*{10}}
     \put (280,30) {\makebox(-1,1)[bl]{$V_2$}}
  \put(275,30) {\line (0,1){50}} 
\put(175,80){\circle*{10}}
     \put (170,80) {\makebox(-1,1)[br]{$V_2$}}
  \put(175,80) {\line (5,3){50}} 
\put(275,80){\circle*{10}}
     \put (280,80) {\makebox(-1,1)[bl]{$V_1$}}
  \put(275,80) {\line (-5,3){50}} 
\put(225,110){\circle*{10}}
     \put (233,110) {\makebox(-1,1)[l]{$D$}}
\put(225,55){\circle*{10}}
     \put (225,65) {\makebox(-1,1)[b]{$V_3$}}
  \put(175,30){\line(2,1){100}}

\put(175,55){\circle*{10}}
     \put (170,55) {\makebox(-1,1)[br]{$V_0$}}
\put(275,55){\circle*{10}}
     \put (280,55) {\makebox(-1,1)[bl]{$V_0$}}

\put (50,-30) {\makebox(-1,1){Net A}}
\put (225,-30) {\makebox(-1,1){Net B}}

\end{picture}

\caption{Hex network}
\label{fig:hex2}
\end{figure}
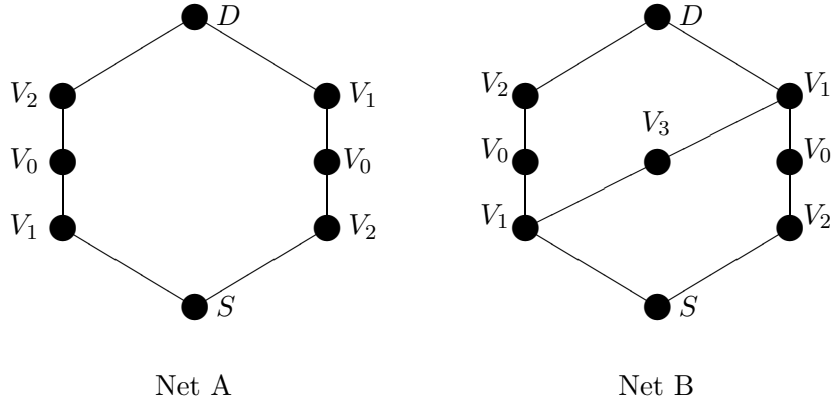

In this section we revisit the network first discussed in 
Section~\ref{sec:ispa} (redrawn in Figure~\ref{fig:hex2} to include
				the dummy nodes). 
In Table~\ref{tab:hex3} we give full results for the load-to-delay
functions discussed in that section. 
We then use load-to-cost functions which are qualitatively
similar to those discussed in Section~\ref{sec:ispa}, but which incorporate 
non-linearities that better represent real router characteristics.
That load-to-cost function and associated results are reported in 
Table~\ref{tab:hex4}.

This network demonstrates that while the addition of a new link may be
beneficial in low traffic cases, it leads to bottlenecks in higher
traffic regimes. 
For ISPA although the per packet cost for loads of 1 and
2 drop drastically when the new link is added, the per packet cost increases
for higher  loads. The MB COIN on the other hand uses the new link 
efficiently. Notice that the MB COIN's performance is slightly worse
than that of the ISPA in the absence of the additional link.
This is caused by the MB COIN having to use a learner to estimate the
WLU values for potential actions whereas the ISPA simply has direct
access to all the information it needs (costs at each link).

\begin{table}[htb]  \centering
\caption{Average Per Packet Cost for HEX network 
for $V_1 = 50 + x \; ; \; V_2 = 10 x \; ; \; V_3 = 10 + x$ .}
%hex3 in sims
\vspace*{.1in}
\begin{tabular}{c|c|c|c} \hline
Load & Net &  ISPA & MB COIN \\ \hline \hline
1 & A &        55.50 &      55.56 \\
  & B &        31.00 &      31.00 \\ \hline
2 & A &        61.00 &      61.10 \\
  & B &        52.00 &      51.69 \\ \hline
3 & A &        66.50 &      66.65 \\
  & B &        73.00 &      64.45 \\ \hline
4 & A &        72.00 &      72.25 \\
  & B &        87.37 &      73.41 \\ \hline
\end{tabular}
\label{tab:hex3}
\end{table}

\begin{table}[htb]  \centering
\caption{Average Per Packet Cost for HEX network 
for $V_1 = 50 + log(1+x) \; ; \; V_2 = 10 x \; ; \; V_3 = log(1+x)$ .}
%hex2 in sims
\vspace*{.1in}
\begin{tabular}{c|c|c|c} \hline
Load & Net &  ISPA & MB COIN \\ \hline \hline
1 & A &        55.41 &      55.44 \\
  & B &        20.69 &      20.69 \\ \hline
2 & A &        60.69 &      60.80 \\
  & B &        41.10 &      41.10 \\ \hline
3 & A &        65.92 &      66.10 \\
  & B &        61.39 &      59.19 \\ \hline
4 & A &        71.10 &      71.41 \\
  & B &        81.61 &      69.88 \\ \hline
\end{tabular}
\label{tab:hex4}
\end{table}

\subsection{Butterfly Network}
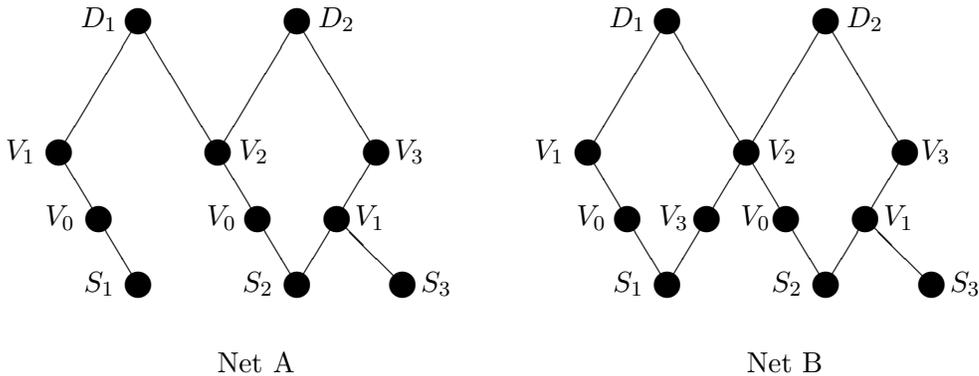
\begin{figure} [htb]
\begin{picture}(200,140)(-40,-30)

\put(30,0){\circle*{10}} 
     \put (22,0) {\makebox(-1,1)[r]{$S_1$}}
  \put(30,0) {\line (-3,5){30}} 
\put(0,50){\circle*{10}}
     \put (-8,50) {\makebox(-1,1)[r]{$V_1$}}
\put(60,50){\circle*{10}}
     \put (68,50) {\makebox(-1,1)[l]{$V_2$}}
  \put(0,50) {\line (3,5){30}} 
  \put(60,50) {\line (-3,5){30}} 
  \put(60,50) {\line (3,5){30}} 
\put(30,100){\circle*{10}}
     \put (23,100) {\makebox(-1,1)[r]{$D_1$}}
\put(90,0){\circle*{10}}
     \put (82,0) {\makebox(-1,1)[r]{$S_2$}}
  \put(90,0) {\line (-3,5){30}}
  \put(90,0) {\line (3,5){30}}
\put(90,100){\circle*{10}}
     \put (98,100) {\makebox(-1,1)[l]{$D_2$}}
\put(120,50){\circle*{10}}
     \put (127,50) {\makebox(-1,1)[l]{$V_3$}}
  \put(120,50) {\line (-3,5){30}}
\put(105,25){\circle*{10}}
     \put (112,25) {\makebox(-1,1)[l]{$V_1$}}
\put(130,0){\circle*{10}}
     \put (137,0) {\makebox(-1,1)[l]{$S_3$}}
  \put(130,0) {\line (-1,1){25}}

\put(15,25){\circle*{10}} 
     \put (7,25) {\makebox(-1,1)[r]{$V_0$}}
\put(75,25){\circle*{10}} 
     \put (68,25) {\makebox(-1,1)[r]{$V_0$}}

\put(230,0){\circle*{10}} 
     \put (222,0) {\makebox(-1,1)[r]{$S_1$}}
  \put(230,0) {\line (-3,5){30}} 
\put(245,25){\circle*{10}} 
     \put (239,25) {\makebox(-1,1)[r]{$V_3$}}
  \put(230,0) {\line (3,5){30}} 
\put(200,50){\circle*{10}}
     \put (192,50) {\makebox(-1,1)[r]{$V_1$}}
\put(260,50){\circle*{10}}
     \put (268,50) {\makebox(-1,1)[l]{$V_2$}}
  \put(200,50) {\line (3,5){30}} 
  \put(260,50) {\line (-3,5){30}} 
  \put(260,50) {\line (3,5){30}} 
\put(230,100){\circle*{10}}
     \put (223,100) {\makebox(-1,1)[r]{$D_1$}}
\put(290,0){\circle*{10}}
     \put (282,0) {\makebox(-1,1)[r]{$S_2$}}
  \put(290,0) {\line (-3,5){30}}
  \put(290,0) {\line (3,5){30}}
\put(290,100){\circle*{10}}
     \put (298,100) {\makebox(-1,1)[l]{$D_2$}}
\put(320,50){\circle*{10}}
     \put (326,50) {\makebox(-1,1)[l]{$V_3$}}
  \put(320,50) {\line (-3,5){30}}
\put(305,25){\circle*{10}}
     \put (312,25) {\makebox(-1,1)[l]{$V_1$}}
\put(330,0){\circle*{10}}
     \put (337,0) {\makebox(-1,1)[l]{$S_3$}}
  \put(330,0) {\line (-1,1){25}}

\put(215,25){\circle*{10}} 
     \put (208,25) {\makebox(-1,1)[r]{$V_0$}}
\put(275,25){\circle*{10}} 
     \put (270,25) {\makebox(-1,1)[r]{$V_0$}}

\put (75,-30) {\makebox(-1,1){Net A}}
\put (275,-30) {\makebox(-1,1){Net B}}

\end{picture}
\caption{Butterfly Network}
\label{fig:butterfly}
\end{figure}

The next network we investigate is shown in Figure~\ref{fig:butterfly}.
It is an extension to the simple network discussed in Section~\ref{sec:bootes}.
We now have doubled the size of the network and have three sources that
have to route their packets to two destinations (packets originating at
$S_1$ go to  $D_1$, and packets originating at $S_2$ 
or $S_3$ go to $D_2$). 
Initially the two halves of the network have minimal contact, but with
the addition of the extra link two sources from the two two halves of the 
network share a common router on their potential shortest path.

Table~\ref{tab:butterfly4} presents two sets of results: first we
present results for uniform traffic through all three sources, and
then results for asymmetric traffic. For the first case, the Braess'
paradox is apparent in the ISPA: adding the new link is beneficial
for the network at low load levels where the average per packet cost
is reduced by nearly $20\%$, but deleterious at higher levels. 
The MB COIN, on the other hand, provides the benefits of the added link 
for the low traffic levels, without suffering from deleterious effects
at higher load levels.

\begin{table}[htb]  \centering
\caption{Average Per Packet Cost for BUTTERFLY network 
for $V_1 = 50 + log(1 + x) \; ; \; V_2 = 10 x \; ; \; V_3 = log(1 + x)$. }
%butterfly4 in sims
\vspace*{.1in}
\begin{tabular}{c|c|c|c} \hline
Loads  $(S_1,S_2,S_3)$  & Net &  ISPA & MB COIN \\ \hline \hline
1,1,1 & A &        112.1 &      112.7 \\
      & B &        92.1  &      92.3 \\ \hline
2,2,2 & A &        123.3 &      124.0 \\
      & B &        133.3 &      122.5 \\ \hline
4,4,4 & A &        144.8 &      142.6 \\
      & B &        156.5 &      142.3 \\ \hline \hline
3,2,1 & A &        81.8  &      82.5 \\
      & B &        99.5  &      81.0 \\ \hline
6,4,2 & A &        96.0  &      94.1 \\
      & B &        105.3 &      94.0 \\ \hline
9,6,3 & A &        105.5 &      98.2 \\
      & B &        106.7 &      98.8 \\ \hline
\end{tabular}
\label{tab:butterfly4}
\end{table}

For the asymmetric traffic patterns, the added link causes a drop
in performance for the ISPA, especially for low overall traffic
levels. This is not true for the MB COIN. Notice also that in the high,
asymmetric traffic regime, the ISPA performs significantly worse than the 
MB COIN
even without the added link, showing that a bottleneck occurs on the right
side of network alone (similar to the Braess' paradox observed in
Section~\ref{sec:bootes}).

\subsection{Ray Network}
\label{sec:ray}
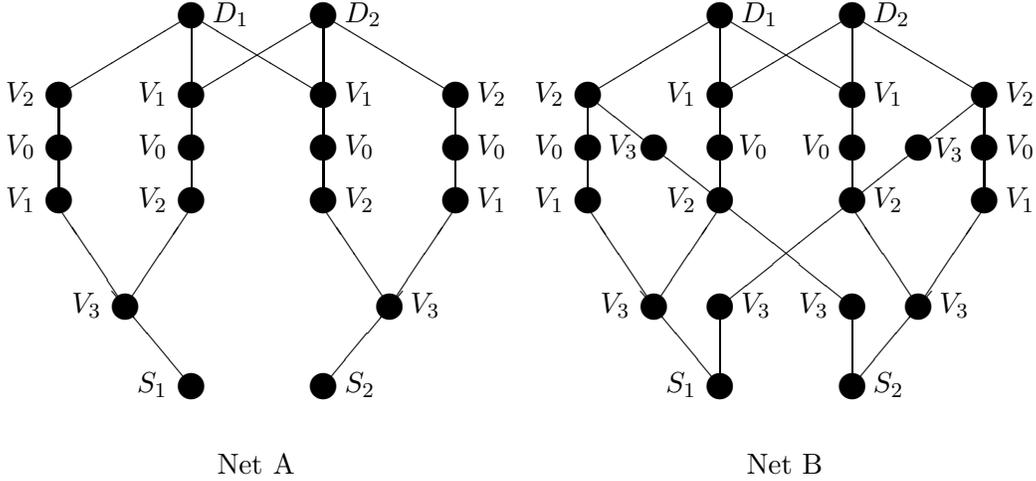
\begin{figure} [htb]
\begin{picture}(200,200)(10,-20)

% first network:
% source layer:
\put(100,0){\circle*{10}} \put (92,0) {\makebox(-1,1)[r]{$S_1$}}
  \put(100,0) {\line (-5,6){30}}
\put(150,0){\circle*{10}} \put (158,0) {\makebox(-1,1)[l]{$S_2$}}
  \put(150,0) {\line (5,6){30}}

% second layer:
\put(75,30){\circle*{10}} \put (67,30) {\makebox(-1,1)[r]{$V_3$}}
  \put(75,30) {\line (-2,3){25}}
  \put(75,30) {\line (2,3){25}}
\put(175,30){\circle*{10}} \put (183,30) {\makebox(-1,1)[l]{$V_3$}}
  \put(175,30) {\line (-2,3){25}}
  \put(175,30) {\line (2,3){25}}

% third layer:
\put(50,70){\circle*{10}} \put (42,70) {\makebox(-1,1)[r]{$V_1$}}
  \put(50,70) {\line (0,1){40}}
\put(100,70){\circle*{10}} \put (92,70) {\makebox(-1,1)[r]{$V_2$}}
  \put(100,70) {\line (0,1){40}}
\put(150,70){\circle*{10}} \put (158,70) {\makebox(-1,1)[l]{$V_2$}}
  \put(150,70) {\line (0,1){40}}
\put(200,70){\circle*{10}} \put (208,70) {\makebox(-1,1)[l]{$V_1$}}
  \put(200,70) {\line (0,1){40}}

% dummy layer:
\put(50,90){\circle*{10}} \put (42,90) {\makebox(-1,1)[r]{$V_0$}}
\put(100,90){\circle*{10}} \put (92,90) {\makebox(-1,1)[r]{$V_0$}}
\put(150,90){\circle*{10}} \put (158,90) {\makebox(-1,1)[l]{$V_0$}}
\put(200,90){\circle*{10}} \put (208,90) {\makebox(-1,1)[l]{$V_0$}}

% fourth layer:
\put(50,110){\circle*{10}} \put (42,110) {\makebox(-1,1)[r]{$V_2$}}
  \put(50,110) {\line (5,3){50}}
\put(100,110){\circle*{10}} \put (92,110) {\makebox(-1,1)[r]{$V_1$}}
  \put(100,110) {\line (0,1){30}}
  \put(100,110) {\line (5,3){50}}
\put(150,110){\circle*{10}} \put (158,110) {\makebox(-1,1)[l]{$V_1$}}
  \put(150,110) {\line (0,1){30}}
  \put(150,110) {\line (-5,3){50}}
\put(200,110){\circle*{10}} \put (208,110) {\makebox(-1,1)[l]{$V_2$}}
  \put(200,110) {\line (-5,3){50}}

% destination layer:
\put(100,140){\circle*{10}} \put (108,140) {\makebox(-1,1)[l]{$D_1$}}
\put(150,140){\circle*{10}} \put (158,140) {\makebox(-1,1)[l]{$D_2$}}

% second network:
% first layer:
\put(300,0){\circle*{10}} \put (292,0) {\makebox(-1,1)[r]{$S_1$}}
  \put(300,0) {\line (-5,6){30}}
  \put(300,0) {\line (0,1){30}}
\put(350,0){\circle*{10}} \put (358,0) {\makebox(-1,1)[l]{$S_2$}}
  \put(350,0) {\line (5,6){30}}
  \put(350,0) {\line (0,1){30}}

% second layer:
\put(275,30){\circle*{10}} \put (267,30) {\makebox(-1,1)[r]{$V_3$}}
  \put(275,30) {\line (-2,3){25}}
  \put(275,30) {\line (2,3){25}}
\put(300,30){\circle*{10}} \put (308,30) {\makebox(-1,1)[l]{$V_3$}}
  \put(300,30) {\line (5,4){50}}
\put(375,30){\circle*{10}} \put (383,30) {\makebox(-1,1)[l]{$V_3$}}
  \put(375,30) {\line (-2,3){25}}
  \put(375,30) {\line (2,3){25}}
\put(350,30){\circle*{10}} \put (342,30) {\makebox(-1,1)[r]{$V_3$}}
  \put(350,30) {\line (-5,4){50}}

% third layer:
\put(250,70){\circle*{10}} \put (242,70) {\makebox(-1,1)[r]{$V_1$}}
  \put(250,70) {\line (0,1){40}}
\put(300,70){\circle*{10}} \put (292,70) {\makebox(-1,1)[r]{$V_2$}}
  \put(300,70) {\line (0,1){40}}
  \put(300,70) {\line (-5,4){50}}
\put(350,70){\circle*{10}} \put (358,70) {\makebox(-1,1)[l]{$V_2$}}
  \put(350,70) {\line (0,1){40}}
  \put(350,70) {\line (5,4){50}}
\put(400,70){\circle*{10}} \put (408,70) {\makebox(-1,1)[l]{$V_1$}}
  \put(400,70) {\line (0,1){40}}

% three and a half layer: 
\put(275,90){\circle*{10}} \put (270,90) {\makebox(-1,1)[r]{$V_3$}}
\put(375,90){\circle*{10}} \put (381,89) {\makebox(-1,1)[l]{$V_3$}}

% dummy layer:
\put(250,90){\circle*{10}} \put (242,90) {\makebox(-1,1)[r]{$V_0$}}
\put(300,90){\circle*{10}} \put (307,90) {\makebox(-1,1)[l]{$V_0$}}
\put(350,90){\circle*{10}} \put (343,90) {\makebox(-1,1)[r]{$V_0$}}
\put(400,90){\circle*{10}} \put (408,90) {\makebox(-1,1)[l]{$V_0$}}

% fourth layer:
\put(250,110){\circle*{10}} \put (242,110) {\makebox(-1,1)[r]{$V_2$}}
  \put(250,110) {\line (5,3){50}}
\put(300,110){\circle*{10}} \put (292,110) {\makebox(-1,1)[r]{$V_1$}}
  \put(300,110) {\line (0,1){30}}
  \put(300,110) {\line (5,3){50}}
\put(350,110){\circle*{10}} \put (358,110) {\makebox(-1,1)[l]{$V_1$}}
  \put(350,110) {\line (0,1){30}}
  \put(350,110) {\line (-5,3){50}}
\put(400,110){\circle*{10}} \put (408,110) {\makebox(-1,1)[l]{$V_2$}}
  \put(400,110) {\line (-5,3){50}}

% destination layer:
\put(300,140){\circle*{10}} \put (308,140) {\makebox(-1,1)[l]{$D_1$}}
\put(350,140){\circle*{10}} \put (358,140) {\makebox(-1,1)[l]{$D_2$}}

%labels
\put (125,-30) {\makebox(-1,1){Net A}}
\put (325,-30) {\makebox(-1,1){Net B}}

\end{picture}
\caption{Ray network}
\label{fig:ray}
\end{figure}

In all the networks and traffic regimes discussed so far the sources are 
the only routers with more than one routing option. 
The final network we investigate is a larger network where the number
of routers with multiply options is significantly higher
than in the previous networks.
Figure~\ref{fig:ray}
shows the initial network (Net A) and the ``augmented'' network (Net B),
where new links have been added. The original network has relatively
few choices for the routers, as packets are directed toward their 
destinations along ``conduits.'' The new links are added in the augmented
networks to provide new choices (crossing patterns) that could be beneficial
if certain of the original conduits experience large costs. 

Table~\ref{tab:ray1} shows the simulation results for these networks
($S_1$ and $S_2$ send packets to $D_1$ and $D_2$ respectively).
At low load levels both the ISPA and the MB COIN use the new links
effectively,
although the MB COIN performs slightly worse. This is mainly caused by
the difficulty encountered by  the simple learner (single nearest neighbor 
algorithm) in quickly learning the traffic patterns in this large network. 
Unlike the ISPA however, the MB COIN avoids the
Braess' paradox in all cases except the very high traffic regime.
Moreover, even there, the effect is significantly milder than that 
encountered by the ISPA.

\begin{table}[htb]  \centering
\caption{Average Per Packet Cost for RAY network 
for $V_1 = 50 + log(1+x) \; ; \; V_2 = 10 x \; ; \; V_3 = 10 + log(1+x)$.}
%big1 in sims
\vspace*{.1in}
\begin{tabular}{c|c|c|c} \hline
Loads at $S_1 and S_2)$ & Net &  ISPA & MB COIN \\ \hline \hline
2,2 & A &        143.6 &      143.7 \\
    & B &        124.4 &      126.9 \\ \hline
3,3 & A &        154.6 &      154.9 \\
    & B &        165.5 &      151.0 \\ \hline
4,4 & A &        165.4 &      166.0 \\
    & B &        197.7 &      165.6 \\ \hline
6,6 & A &        186.7 &      187.4 \\
    & B &        205.1 &      191.6 \\ \hline
\end{tabular}
\label{tab:ray1}
\end{table}

\subsection{Steering the MB COIN}
\label{sec:steer}
\begin{figure}[bth]
\centering
\subfigure[Hex4] {\label{fig:mbhex4}
\includegraphics[width=2.8in,height=2.2in]{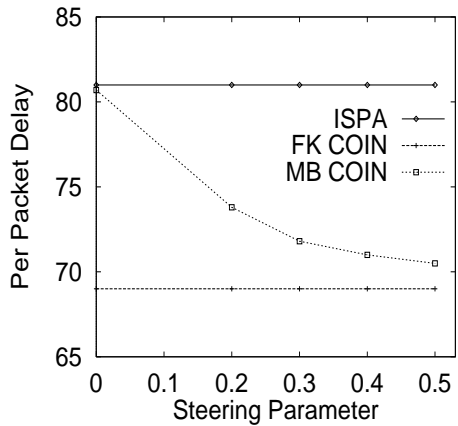}}
\subfigure[Ray4]{\label{fig:mbray4}
\includegraphics[width=2.8in,height=2.2in]{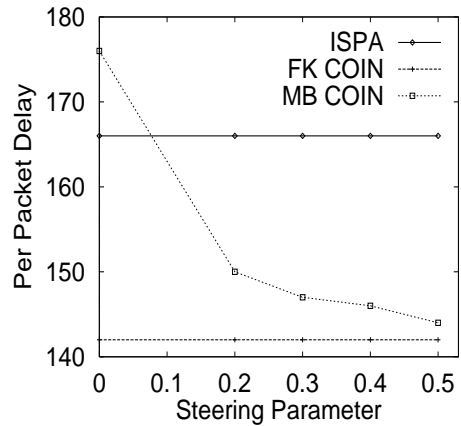}}
\subfigure[Butterfly4]{\label{fig:mbbut4}
\includegraphics[width=2.8in,height=2.2in]{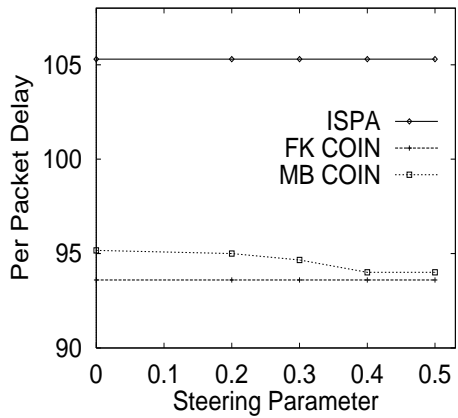}}
\subfigure[Bootes4]{\label{fig:mbboo4}
\includegraphics[width=2.8in,height=2.2in]{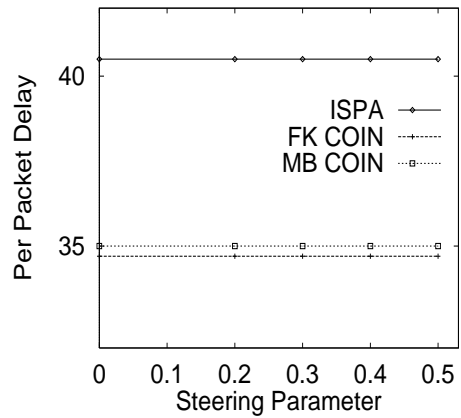}}
\caption{Impact of switching.}
\label{fig:mbswitch}
\end{figure}

The final aspect of COIN-based routing we investigate is the
impact of the choice for the value of the steering parameter.
This parameter both controls the amount of exploration the algorithm 
performs and determines the ``intelligence'' of the MB COIN at
estimating the surface directly calculated by the FK COIN. 
In Figure~\ref{fig:mbswitch} we provide results
when the steering parameter is set to $1.0$ so that the MB COIN reduces 
to FK COIN. This 
provides an upper bound on the performance that the MB COIN could achieve
if it used no exploration.

For the HEX network (Figure~\ref{fig:mbhex4}), the performance at the 
worst setting for the MB COIN, which corresponds to no steering,
is comparable to ISPA. In contrast, with moderate steering (0.5) 
the results are similar to that of the FK COIN,
as the learner has more information to work with
(arising from the extra parts of the input space represented in 
  the training set due to the occasional use of the FK COIN), 
it bridges the gap between a suboptimal algorithm susceptible
to Braess' paradox and one which efficiently avoids that paradox. 

For the RAY network (Figure~\ref{fig:mbray4}), the value of the steering
parameter is more critical. 
With no steering at all, the MB COIN
performs poorly in this network --- even worse than ISPA. This is not 
surprising in that because there are many routing choices that affect the 
performance, the simple memory-based learner needs
proper ``seeding'' to be able to perform well. In any case, with the addition
of steering the MB COIN quickly outperforms the ISPA. 

Finally, for both the Butterfly and Bootes networks 
(Figures~\ref{fig:mbbut4}~-~\ref{fig:mbboo4}) the MB COIN needs very 
little steering to perform well. Although for the Butterfly network the 
performance of MB COIN improves slightly with more information, 
it is significantly better than the ISPA across the board.

\section{CONCLUSION}
Effective routing in a network is a fundamental problem in many fields,
including data communications and transportation.
Shortest path algorithms provide an elegant solution to this 
problem, but under certain circumstances suffer from less than 
desirable effects.
One such effect is Braess' paradox, where increased capacity results
in lower overall throughput for shortest path algorithms due to the
potentially harmful side-effects of the decisions made by such 
algorithms. Even a full-blown load-balancing can suffer from such
side-effects , since in general they extend across time as well as space, 
whereas, load balancing ignores temporal side-effects.

Collective Intelligence is a novel way of controlling distributed
systems so as to avoid the deleterious side-effects. 
In a COIN, the central issue is in determining
the personal objectives to be assigned to the components of the system. 
One wants to choose those goals so that the greedy
pursuit of those goals by the components of the system leads to a globally desirable solution. We
have summarized COIN theory and derived a routing algorithm based
on that theory. 
In our simulations, the ISPA induced  average costs as much as 32 \%
higher than the COIN-based algorithm. This was despite the ISPA's having
access to more information than the MB COIN. Furthermore the COIN-based
algorithm avoided the Braess' paradoxes that seriously diminished
the performance of the ISPA.

In the work presented here, the COIN-based algorithm had to overcome
severe limitations. Firstly, the estimation of the effect sets 
used were exceedingly poor.  Secondly, the learners 
were particularly simple-minded, and therefore were not able
to effectively maximize their performance. That a COIN-based router with such
serious limitations consistently outperformed an ideal shortest path algorithm
demonstrates the strength of the proposed method. 

We are currently
investigating novel utilities that are more ``learnable'' for the
routers as well as expand the simulations to larger networks using
a commercial event driven simulator.
Future work will focus on not making the approximation that current 
traffic levels do not affect future windowed loads (Equation~\ref{eq:netwlu}).
It will also involve
investigating better estimates of effect sets, in particular not
including all nodes with the same destination in one's effect set, and more
generally using a more "fine-grained" representation of the nodes, for
example including each packet's originating source, to allow a more
fine-grained effect set (and resultant WLU).

\section{Acknowledgements}

We would like to thank Joe Sill for helpful discussion.

\bibliographystyle{plain}

\end{document}